\documentclass[aip,rsi,amsmath,amssymb,reprint]{revtex4-2}

\usepackage{graphicx}% Include figure files
\usepackage{dcolumn}% Align table columns on decimal point
\usepackage{bm}% bold math

\usepackage[utf8]{inputenc}
\usepackage[T1]{fontenc}
\usepackage{mathptmx}
\usepackage{mathtools}
\usepackage{etoolbox}
\usepackage{color}
\usepackage{hyperref}
\hypersetup{
  colorlinks,
  citecolor=blue,
  linkcolor=blue,
  urlcolor=blue}

\newcommand{\BS}{Bi$_2$Se$_3$}

\newcommand{\ie}{{\em i.e.}}

\newcommand{\cf}{{\em cf.}}
\newcommand{\rbf}{\mathbf{r}}
\newcommand{\vbf}{\mathbf{v}}
\newcommand{\Bbf}{\mathbf{B}}
\newcommand{\Ebf}{\mathbf{E}}

\makeatletter

\def\@allemails{}

\def\mark@corresponding#1#2{%
  \def\@author{{#1\textsuperscript{*}}{#2}}%
}

\def\@email#1#2{%
  \endgroup
  \expandafter\mark@corresponding\@author
  \ifx\@allemails\@empty
    \gdef\@allemails{\href{mailto:#2}{#2}}%
  \else
    \gappto\@allemails{, \href{mailto:#2}{#2}}%
  \fi
}

\AtBeginDocument{%
  \patchcmd{\titleblock@produce}
    {\frontmatter@RRAPformat}
    {\frontmatter@RRAPformat
      {\produce@RRAP{*Electronic mail: \@allemails}}%
     \frontmatter@RRAPformat}
    {}{}%
}

\makeatother
\begin{document}

\preprint{AIP/123-QED}

\title{High-resolution angle-resolved photoemission spectroscopy with tunable magnetic field}

% ***** Authors
\author{Jairo Obando-Guevara}
\email{jairo.obando@dipc.org}
\thanks{These two authors contributed equally}
\affiliation{ 
Department of Physics, The Pennsylvania State University, University Park, Pennsylvania, 16802 USA
}%

\author{Huu-Thong Le}
\thanks{These two authors contributed equally}
\affiliation{ 
Department of Physics, The Pennsylvania State University, University Park, Pennsylvania, 16802 USA
}%

\author{Emiliano Corcino Aracena}
\affiliation{ 
Department of Physics, The Pennsylvania State University, University Park, Pennsylvania, 16802 USA
}%

\author{Sayan Singha}
\affiliation{ 
Department of Physics, The Pennsylvania State University, University Park, Pennsylvania, 16802 USA
}%

\author{Kyungchan Lee}
\affiliation{ 
Department of Physics, The Pennsylvania State University, University Park, Pennsylvania, 16802 USA
}%

\author{Yingdong Guan}
\affiliation{ 
Department of Physics, The Pennsylvania State University, University Park, Pennsylvania, 16802 USA
}%

\author{Zhiqiang Mao}
\affiliation{ 
Department of Physics, The Pennsylvania State University, University Park, Pennsylvania, 16802 USA
}%

\author{Dezhe Jin}
\affiliation{ 
Department of Physics, The Pennsylvania State University, University Park, Pennsylvania, 16802 USA
}%
\affiliation{Center for Theory of Emergent Quantum Matter, The Pennsylvania State University, University Park, Pennsylvania 16802, USA}

\author{Chaoxing Liu}
\affiliation{ 
Department of Physics, The Pennsylvania State University, University Park, Pennsylvania, 16802 USA
}%
\affiliation{Center for Theory of Emergent Quantum Matter, The Pennsylvania State University, University Park, Pennsylvania 16802, USA}

\author{Heike Pfau}
\email{heike.pfau@psu.edu}
\affiliation{ 
Department of Physics, The Pennsylvania State University, University Park, Pennsylvania, 16802 USA
}%

\date{\today}

%------------------------------------------------------------
% Abstract
%------------------------------------------------------------
\begin{abstract}
The control and perturbation of quantum phenomena with magnetic fields is an indispensable tool in materials research. Recently, in-situ field tuning was implemented into angle-resolved photoemission spectroscopy (magneto-ARPES), which provides direct momentum and energy-resolved information of the field-dependent electronic structure. However, aberrations of the electron trajectories were shown to be substantial, leading to significant spectral distortions and broadening even in small fields. Here we show that the electronic structure can be recovered from magneto-ARPES spectra with high accuracy while maintaining high momentum resolution even when strong trajectory aberrations are present. We studied \BS~in a dipole field of a coil using a laser-based ARPES system. The electronic structure was reconstructed in post-processing using detailed electron trajectory simulations. We identify two-dimensional (2D) momentum mapping, a micron beam spot size, and precise numerical field simulations as critical technical requirements for high-resolution magneto-ARPES. We show how circular dichroism can provide additional information about the coupling of the magnetic field to the spin. The experimental achievements and the scaling laws from our simulations provide a road map towards magneto-ARPES in larger fields.
\end{abstract}

\maketitle

% -----------------------------------------------------------
% Intro
% -----------------------------------------------------------

%%%%%%%%%%%%%%%%%%%%%%%%%%%%%%%%%%%%%%%%%%%%%%%%%%%%%%%%%%%%%%
\section{Introduction}
%%%%%%%%%%%%%%%%%%%%%%%%%%%%%%%%%%%%%%%%%%%%%%%%%%%%%%%%%%%%%%

Angle-resolved photoemission spectroscopy (ARPES) is a prominent technique in materials research. It measures the energy and momentum-resolved single-particle spectral function of the occupied electronic states. Photoemission matrix elements under different light polarization provide additional orbital resolution, coupling with spin polarimeters gives spin information, and pump-probe detection schemes access unoccupied states and electron dynamics. Therefore, ARPES delivers essential information on the electronic many-body wave functions and plays a central role in the discovery and understanding of quantum materials. \cite{Damascelli_2004,sobota_2021,boschini_2024}

Nonthermal tuning of quantum materials is an indispensable tool to uncover emergent phenomena and the fundamental interactions that drive them. In-situ perturbation by pressure, electric and magnetic fields is of particular importance because they circumvent disorder effects induced by chemical doping and substitution. Recent technical advances now enable ARPES with in-situ stress tuning \cite{Jo_2024_review,Ricco_2018,pfau_2019,cai_2020,Zhang_2021,Nicholson_2021,hyun_2022,jo_2024}, gating \cite{Joucken_2019,jones_2020,Nguyen_2019}, and current \cite{kaminski_2016}.

Recent pioneering work added ARPES under in-situ tunable magnetic field (magneto-ARPES). They implemented the field using a solenoid around the sample \cite{huang_2023,huang_2026} or by changing the magnetization of a ferromagnetic yoke \cite{ryu_2023}. A magnetic field distorts the trajectories of the photoemitted electrons from the sample to the ARPES analyzer. The aberrations were shown to be significant even for small fields in the milli-Tesla range. This led to substantial spectral distortions and broadening of the ARPES spectra. Experiments on magnetic substrates demonstrated that higher multi-pole field configurations can reduce trajectory distortions.\cite{Li_2025} The nano-structured magnetic films limit measurements to thin films in a fixed field. 

To expand magneto-ARPES to a broad range of materials and quantum phenomena, it is imperative to demonstrate that the spectral function can be accurately determined in a variable field while maintaining the high resolution of modern instruments. Here, we present magneto-ARPES measurements on \BS~using a custom, high-resolution laser-based instrument, which is optimized for the application of in-situ tuning parameters. A dipole field is provided by a solenoid around the sample. Combining experiments with detailed finite-element analysis (FEA), we demonstrate that the electronic structure can be recovered with high accuracy in post-processing even when strong trajectory aberrations are present. Furthermore, we can maintain high momentum resolution over the whole accessible field range. Two-dimensional (2D) momentum mapping and precise characterization of the field with FEA are indispensable for trajectory corrections. A small beam spot size is crucial for high resolution. Scaling laws from our simulations provide a road map for the implementation and expansion of magneto-ARPES in different instruments and with increasing field range. In addition to field-dependent dispersions, we showcase how circular dichroism can probe the coupling of the magnetic field to spin degrees of freedom in materials with spin-orbit coupling.

%%%%%%%%%%%%%%%%%%%%%%%%%%%%%%%%%%%%%%%%%%%%%%%%%%%%%%%%%%%%%%
\section{Experimental Setup}
%%%%%%%%%%%%%%%%%%%%%%%%%%%%%%%%%%%%%%%%%%%%%%%%%%%%%%%%%%%%%%

% -----------------------------------------------------------
% Figure: Sketch and Simulation of Coil
% -----------------------------------------------------------
\begin{figure}
\includegraphics[width=1\columnwidth]{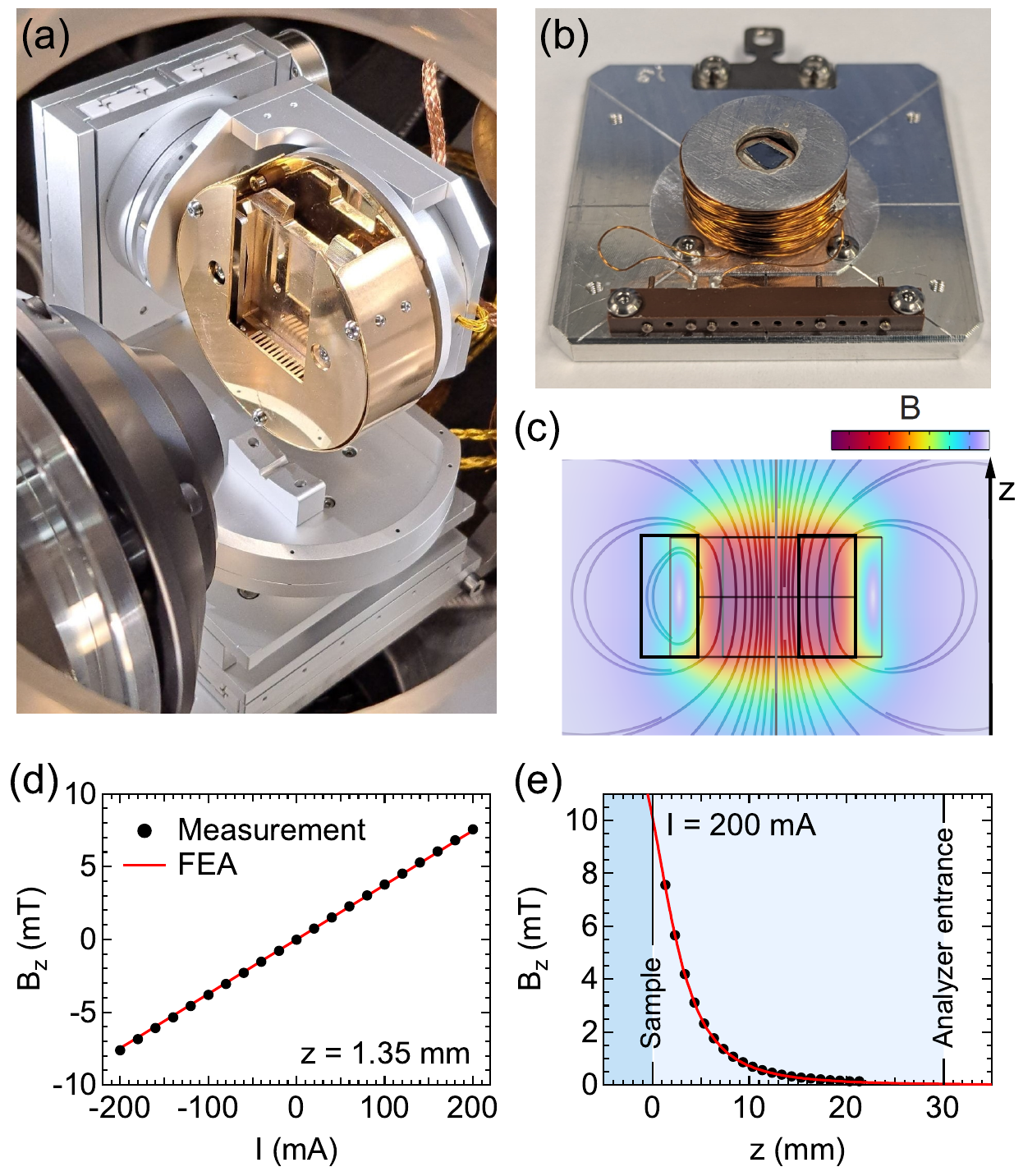}
\caption{
Magneto-ARPES setup. (a) 6-axis piezo-driven manipulator with a sample receiver that is equipped with 12 electrical contacts. (b) Sample holder with solenoid. \BS~sample is mounted in the center of the coil. (c) FEA of the magnetic field of the solenoid shown on a cross section through the coil. (d,e) Comparison of the measured and FEA result for the $z$-component of the magnetic field $B_z$ as a function of current and distance $z$. The sample is located at $z=0$\,mm. $z=1.35$\,mm is slightly above the solenoid.  
}
\label{Fig:Coil}
\end{figure}

An in-situ tunable magnetic field is provided by a coil of Kapton-insulated copper wire (Fig.~\ref{Fig:Coil}(b)) with dimensions as listed in Table \ref{Tab:coils}. We characterized the magnetic field using a Hall sensor (Fig.~\ref{Fig:Coil}(d,e)). The coil provides a maximum field at the sample position of 10\,mT within the current limitations of the copper wire. The coil resistance is 2.5~$\Omega$ at 30\,K, corresponding to a Joule heating power of $P=I^2R=0.1$\,W at maximum current. This leads to a 4\,K temperature increase in our system.

% -----------------------------------------------------------
% Table Coil parameters
% -----------------------------------------------------------
    \begin{table}[h]
        \centering
        \begin{tabular}{c c c c c}
            \hline\hline
                $r_i$ (mm) & $r_o$ (mm) & $h$ (mm) & $N$ & $A$ ($\text{mm}^2$)\\
            \hline
            %Coil 2 & 3.17 & 6.5 & 7 & 686 & 0.02\\
            %Coil 4 & 3.17 & 6.3 & 7.13 & 617 & 0.02\\
            3.2 & 6.4 & 7 & 650 & 0.02\\
            \hline\hline
        \end{tabular}
        \caption{Summary of the coil dimensions: inner radius $r_i$, outer radius $r_o$, height $h$, number of windings $N$, and wire cross section $A$.}
        \label{Tab:coils}
    \end{table}

The large sample holder (35\,mm x 35\,mm) is equipped with 12 electrical contacts, which match contact springs at the sample receiver on the vacuum side (Fig.~\ref{Fig:Coil}(a)). The sample receiver is mounted on a 6-axis piezo-driven manipulator for sample positioning with a precision and a bi-directional repeatability on the nanometer scale. The ARPES system is paired with a 6\,eV continuous-wave laser, which minimizes space charge. Our optical setup allows for full polarization control and focuses the light onto a 20\,$\mu$m spot on the sample. In combination with the precision sample manipulator, we are able to perform measurements in the micro-ARPES regime. We will show below that this is crucial to maintain momentum resolution in magneto-ARPES.

The system is equipped with a MB Scientific A-1 hemispherical electron analyzer that features 2D momentum mapping, which provides access to both in-plane momentum directions without sample rotation. We achieve an energy and momentum resolution of 1 meV and 0.001 \AA$^{-1}$. The analyzer slit is oriented horizontally in the same plane as the light incident angle. The analyzer can be biased, which increases the electron kinetic energy and therefore expands momentum access within a single spectrum to the full photoemission cone \cite{gauthier_2021,Miao_2026}. The higher kinetic energy also reduces trajectory distortions in magnetic fields as we show below.

The compact manipulator design allows for a small mu-metal vacuum chamber optimized for pumping geometry. We reach a pressure of $7\cdot10^{-12}$\,torr at room temperature as measured by an extraction gauge. The sample can be cooled down by a liquid helium flow cryostat.

%%%%%%%%%%%%%%%%%%%%%%%%%%%%%%%%%%%%%%%%%%%%%%%%%%%%%%%%%%%%%%
\section{Electron Trajectory Simulations}
%%%%%%%%%%%%%%%%%%%%%%%%%%%%%%%%%%%%%%%%%%%%%%%%%%%%%%%%%%%%%%

% -----------------------------------------------------------
% Figure: Trajectory simulations
% -----------------------------------------------------------
\begin{figure*}
\includegraphics[width=\textwidth]{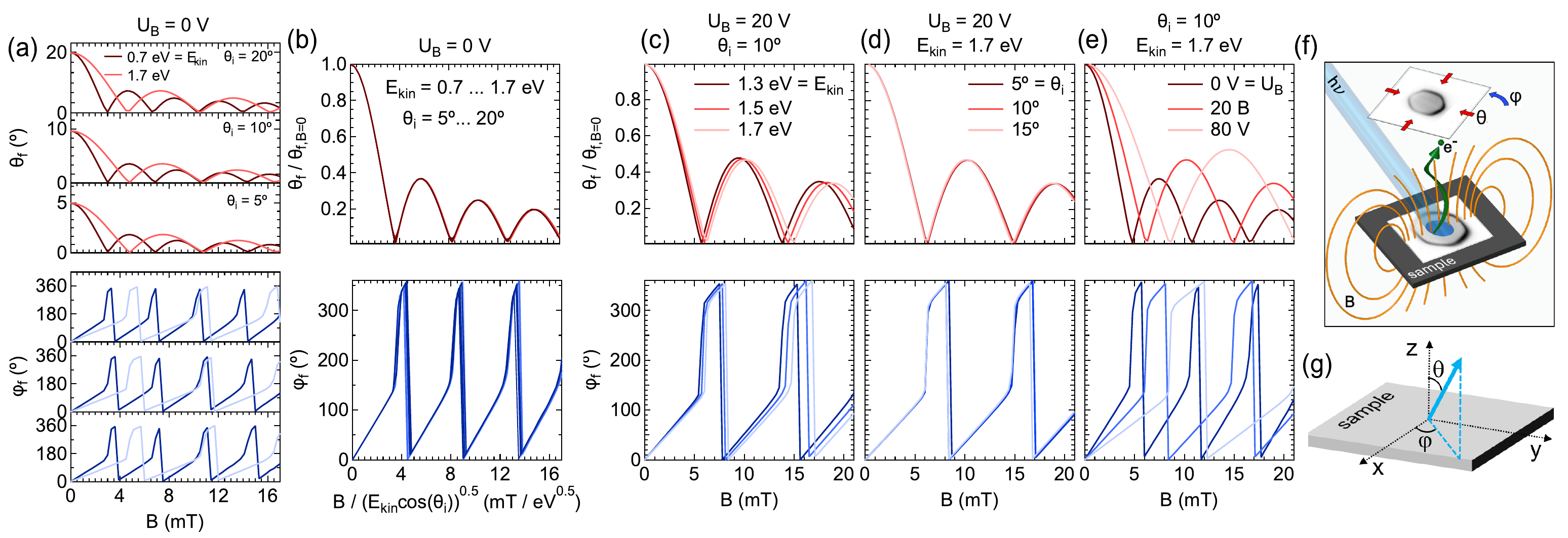}
\caption{
Electron trajectory simulations. (a) Contraction angle $\theta_f$ (top) and rotation angle $\varphi_f$ (bottom) at the analyzer entrance as a function of field $B$ at the sample position. We used initial $\varphi_i=0$, zero bias voltage $U_\mathrm{B}$ and a range of initial kinetic energies of the photoemitted electron $E_\mathrm{kin}$ and photoemission angles $\theta_i$. The same parameters were used for $\theta_f$ and $\varphi_f$. (b) Corresponding scaling plot for (a). Jumps in $\varphi_f$ reflect a change of the geometric quadrant when $\varphi_f= 360^{\circ}$ ($0^\circ\le\varphi_f<360^\circ$) or $\theta_f= 0^{\circ}$ (focal points). If $\theta_f= 0^{\circ}$, trajectories are straight and ARPES spectra are compressed to a point. (c--e) Simulations with non-zero bias voltage as a function of (c) $E_\mathrm{kin}$, (d) $\theta_i$, and (e) $U_\mathrm{B}$. (f) Illustration of rotation and contraction of Fermi surface images due to the dipole field. (g) Angle definitions with $0^\circ\le\varphi<360^\circ$ and $0^\circ\le\theta<90^\circ$.
}
\label{Fig:simulations}
\end{figure*}

The magnetic field $\bf B$ of the solenoid distorts the trajectories of the photoemitted electrons traveling from the sample to the ARPES analyzer. We model the electron trajectories as a function of solenoid current $I$, electron kinetic energy $E_\mathrm{kin}$, initial photoemission angle at the sample $\theta_i$, and bias $U_\mathrm{B}$ in great detail. The results are used to numerically reconstruct the initial photoemission angles from the measured ARPES spectra in post-processing and obtain the electronic band structure of materials under magnetic fields.

We model the magnetic field by FEA using the experimentally determined coil geometry in Tab.~\ref{Tab:coils}. Figure \ref{Fig:Coil}(c) shows the field magnitude for a slice through the center of the solenoid, depicting the expected dipole geometry. The simulated $B_z$ matches our measurements (Fig.~\ref{Fig:Coil}(d,e)), which confirms that the coil provides a well-defined dipole field captured by our FEA. $B_z$ attenuates by 99.6\% from the sample to the analyzer entrance. We therefore neglect any residual field inside the analyzer lens column in our simulations and in the analysis of our magneto-ARPES data. The bias voltage between sample and analyzer is modeled as a homogeneous electric field $\bf E$. 

To simulate the electron trajectories, we numerically solve the electron equations of motion in this spatially varying electro-magnetic field using the Runge-Kutta-2(3) algorithm \cite{bogacki19893}
\begin{align}
	&
	\begin{dcases}
		\dot{\rbf} &= \vbf,
		\\
		m \dot{\vbf} &= q \Ebf + q \vbf \times \Bbf(\rbf),
	\end{dcases}
\end{align}
where $\rbf, \vbf$ are the electron's position and velocity vectors to be solved for. 

We characterize the electron trajectories in Fig.~\ref{Fig:simulations}(a--e) within a spherical coordinate system (Fig.~\ref{Fig:simulations}(g)) using the angles $\theta_f$ and $\varphi_f$ at the analyzer entrance as a function of the field $B$ at the sample. Generally, a dipole field contracts ($\theta_f$) and rotates ($\varphi_f$) ARPES spectra due to out-of-plane and in-plane field components, respectively, in agreement with previous studies \cite{huang_2023}. The magnitude of these effects depends on $B$, $E_\mathrm{kin}$, $\theta_i$, and $U_\mathrm{B}$. Simulations for zero bias voltage in Fig.~\ref{Fig:simulations}(a,b) reveal a scaling relation 
\begin{align}
\begin{dcases}
    \theta_f/\theta_{f,B=0} &= T(\chi) \\
    \varphi_f &= F(\chi)
\end{dcases}
,\quad
\chi &= \frac{B}{\sqrt{E_\mathrm{kin}\cos\theta_i}}
\label{eq:scaling}
\end{align}
of a single variable $\chi$ within the simulated range of $E_\mathrm{kin}$ and $\theta_i$. All curves collapse onto the scaling functions $F(\chi)$ and $T(\chi)$, which only depend on the geometry of the coil. Trajectories are qualitatively similar once $U_\mathrm{B}\neq0$ (Fig.~\ref{Fig:simulations}(c--e)). However, the scaling fails even when $U_\mathrm{B}$ is added into the kinetic energy term of the scaling variable $\chi$. Reconstruction of the spectral function from our magneto-ARPES data therefore requires detailed characterization of the trajectories over the whole parameter space of the measurement.

% -----------------------------------------------------------
% Figure: Trajectories
% -----------------------------------------------------------
\begin{figure}
\includegraphics[width=\columnwidth]{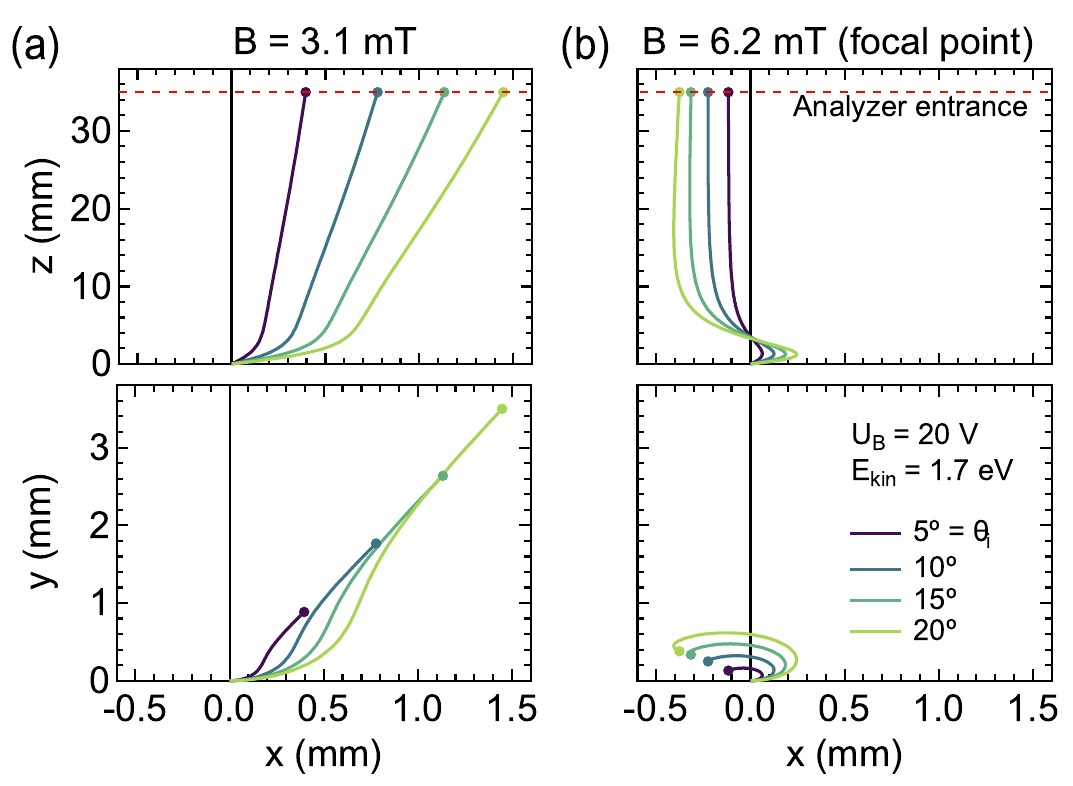}
\caption{
Angular focusing of photoelectrons. Simulated trajectories from the sample ($x,y,z=0$) to the analyzer entrance (red dashed line) for different initial photoemission angle $\theta_i$. (a) $B=3.1$\,mT. (b) $B=6.2$\,mT, \ie~at the first angular focusing point (\cf~Fig.~\ref{Fig:simulations}(d)). The upper (lower) panels show the projection onto the $x-z$ ($x-y$) plane.
}
\label{Fig:trajectories}
\end{figure}

The in-plane rotation of ARPES spectra can be captured with 2D momentum mapping capabilities. The contraction of the spectra has the most significant effect on imaging performance. $\theta_f$ becomes zero at certain field values, \ie~ARPES spectra are compressed to a point.
Fig.~\ref{Fig:trajectories} compares electron trajectories for a field below and at such an angular focus point. Fig.~\ref{Fig:trajectories}(a) highlights the compression from $\theta_i$ at the sample to the smaller $\theta_f$ at the analyzer entrance. At the focal point (Fig.~\ref{Fig:trajectories}(b)), all trajectories are parallel to $z$ at the analyzer entrance and angle resolution is lost. Small deviations to the focusing properties are seen once $\theta_i>15^\circ$. The dipole magnetic field effectively acts as an electron lens with a field-dependent focal length. Importantly, $\theta_f$ recovers to non-zero values in between these focal points, enabling ARPES measurements at higher fields. Larger kinetic energies reduce contraction and shift the focal points to higher fields. However, different light sources provide only a moderate quantitative advantage. For example, 20\,eV kinetic energy in our coil geometry pushes the focal point from 5\,mT to 17\,mT. A bias voltage $U_\mathrm{B}$ achieves similar effects but on a smaller scale, because trajectories are most severely distorted close to the sample.

% -----------------------------------------------------------
% Figure: Broadening
% -----------------------------------------------------------
\begin{figure}
\includegraphics[width=\columnwidth]{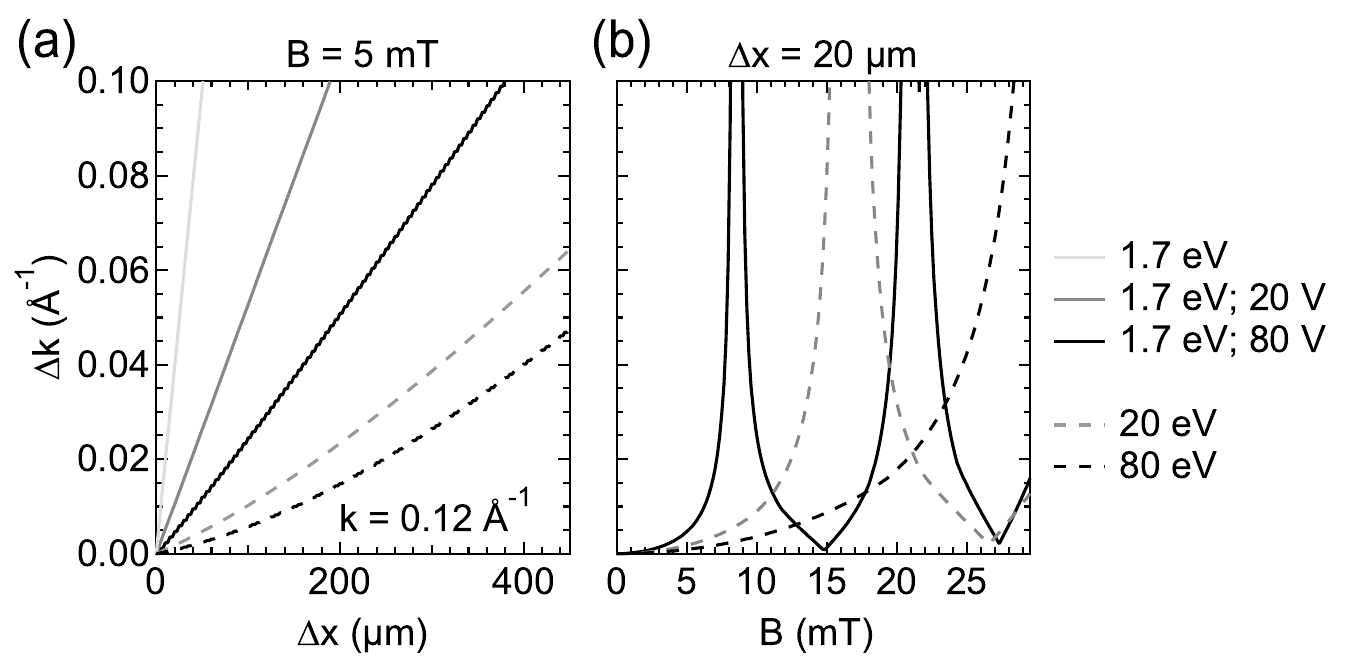}
\caption{
Simulated momentum broadening $\Delta k$ as a function of (a) spot size $\Delta x$ and (b) field $B$ evaluated at a momentum of $k=0.12$\,\AA$^{-1}$ for different kinetic energies and bias voltages. Divergences of $\Delta k$ seen in (b) coincide with the focal points in $\theta_f$.
}
\label{Fig:broadening}
\end{figure}

A non-zero beam spot size $\Delta x$ leads to angle broadening, because electrons photoemitted from a different part of the sample experience a different magnetic field. Fig.~\ref{Fig:broadening}(a) shows an example for photoemission from a fixed momentum of 0.12\,\AA$^{-1}$ close to $k_\mathrm{F}$ of \BS. We evaluate the broadening $\Delta k$ by calculating the maximum deviation $\theta_f(x) - \theta_f(0)$ for the emission location $x\in [-\Delta x/2,+\Delta x/2]$ along a line with $\varphi_i=0$. $\Delta k$ reaches values on the order of the momentum itself for a few hundred micrometer spot sizes, even with bias voltage or higher photon energies. Momentum information is essentially lost in this regime. High-resolution magneto-ARPES can therefore only be accomplished with micro-ARPES setups. Broadening becomes unsustainable at fields close to the focal points independent of kinetic energy due to the strong angle contraction (Fig.~\ref{Fig:broadening}(b)). Importantly, $\Delta k$ has a pronounced minimum in between the focal points, which will allow magneto-ARPES at larger fields. 

%%%%%%%%%%%%%%%%%%%%%%%%%%%%%%%%%%%%%%%%%%%%%%%%%%%%%%%%%%%%%%
\section{Magneto-ARPES on B\lowercase{i}$_2$S\lowercase{e}$_3$}
%%%%%%%%%%%%%%%%%%%%%%%%%%%%%%%%%%%%%%%%%%%%%%%%%%%%%%%%%%%%%%

% -----------------------------------------------------------
% Figure: ARPES
% -----------------------------------------------------------
\begin{figure*}
\includegraphics[width=\textwidth]{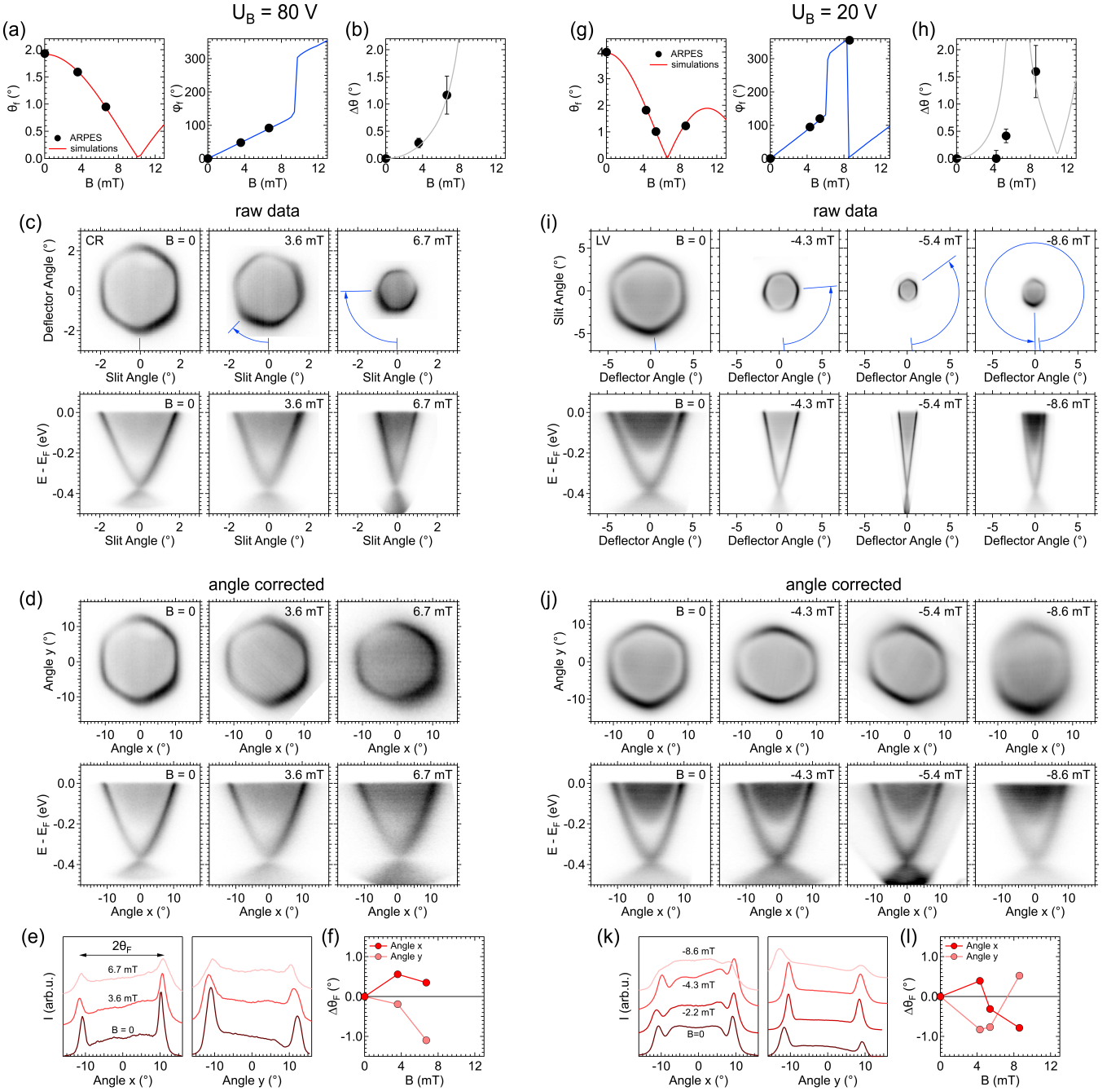}
\caption{
Magneto-ARPES of \BS. (a) Magnetic-field dependence of the Fermi surface contraction angle $\theta_f$ and rotation angle $\varphi_f$ ($E_\mathrm{kin}=1.7\,$\,eV, $\theta_i=11^\circ$, $U_\mathrm{B}=80$\,V), comparing simulations (lines) with ARPES measurements (dots) extracted from data in (c). (b) Magnetic-field-induced angle broadening determined at $k_\mathrm{F}$. The intrinsic line width at zero field is $\Delta\theta_0 = 0.8^\circ$. (c) ARPES Fermi surface maps and energy-angle cuts obtained with CR polarized light using 2D momentum mapping with a bias voltage of $U_\mathrm{B} = 80$\,V. (d) Fermi surface maps and energy-angle cuts after correction using our trajectory simulations. (e) MDCs across the Fermi surface in (d). (f) Field dependence of Fermi surface size with respect to $B=0$ along the $x$ and the $y$ direction obtained from MDCs in (e). (g--l) Same as (a--f) on a second sample measured with $U_\mathrm{B} = 20$\,V and LV polarized light.
}
\label{Fig:ARPES}
\end{figure*}

We performed magneto-ARPES measurements on the topological insulator \BS. It is ideally suited to test the performance of our magneto-ARPES setup. The well-characterized band structure \cite{Zhang_2009,Xia_2009,Cao_2013,Jozwiak_2011} consists of topological Dirac surface states inside a bulk band gap. The effective Hamiltonian of the surface states is given by
\begin{equation}
    H(k) = \alpha (k_x \sigma_y - k_y \sigma_x) + \gamma (k_+^3 + k_-^3) \sigma_z - g \mu_\mathrm{B} B \sigma_z
\label{eqn:H}
\end{equation}
where $\sigma_{x,y,z}$ are spin Pauli matrices, and $k_{\pm} = k_x \pm ik_y$. $\alpha$ is the Fermi velocity and this term defines the in-plane spin-momentum locking, giving rise to a helical spin texture around the Dirac point. The $\gamma$ term introduces a warping and an out-of-plane spin at larger momenta, \ie~close to the Fermi level. The last term describes Zeeman spin splitting in a magnetic field, which mainly influences the band structure close to the Dirac point in an energy range of $g \mu_B B$. The effect of the magnetic field on the remainder of the topological surface states, in particular the Fermi surface, is negligible.

From an experimental point of view, \BS~is an excellent candidate as well. The crystals used in this study were grown with a modified Bridgman technique, are electron-doped \cite{guan_2025}, and provide ARPES data with consistently high spectral quality. The complete Fermi surface can be accessed with 6\,eV photoemission despite the comparatively high work function of approximately 5.5\,eV.

Figure \ref{Fig:ARPES} summarizes our ARPES results on \BS~obtained at 30 K. We present measurements on two samples: the first was measured with $U_\mathrm{B}=80$ V bias voltage and right-handed circularly (CR) polarized light (Fig.~\ref{Fig:ARPES}(a--f)), the second with $U_\mathrm{B}=20$ V bias voltage and linear vertical (LV) polarization (Fig.~\ref{Fig:ARPES}(g--l)). The 2D mapping mode with bias voltage allows us to measure the full photoemission cone along energy, slit angle, and deflector angle. The spectra in Fig.~\ref{Fig:ARPES} show slices through these ARPES cubes.

The zero-field data on both samples show the expected hexagonally warped Fermi surface of the topological surface states and a Dirac point at -0.38\,eV (Fig.~\ref{Fig:ARPES}(c,i)). In addition, data taken with LV polarized light highlight the bulk band in the energy range -0.2 to 0\,eV (Fig.~\ref{Fig:ARPES}(i)). The applied bias leads to an angle compression similar to previous studies \cite{gauthier_2021,Miao_2026}: While $k_\mathrm{F}$ is located at $11^\circ$ at $U_\mathrm{B}=0$ (not shown), it reduces to $2^\circ$ ($4^\circ$) for $U_\mathrm{B}=80$\,V (20\,V) (Fig.~\ref{Fig:ARPES}(c,i)). It was shown that the measured angle can depend on both the angle $\theta_f$ and the position $x_f$ of the electron at the analyzer entrance \cite{Gauthier_2026}. This situation is realized in our setup and we therefore scale the simulations of $\theta_f(B)$ to match experimental results at $B=0$.

ARPES spectra in field show the effects of angle rotation and angle contraction (Fig.~\ref{Fig:ARPES}(c,i)). These effects follow the predictions from our trajectory simulations as shown for $k_\mathrm{F}$ in Fig.~\ref{Fig:ARPES}(a,g). A larger bias voltage reduces both angle rotation and angle contraction. 

To correct for trajectory distortions and reconstruct the electronic structure in field, we numerically map the measured angles $\theta_f$ and $\varphi_f$ onto the initial photoemission angles at the sample surface $\theta_i$, $\varphi_i$ by reversing our trajectory simulations
\begin{eqnarray}
    \theta_f(B,U_\mathrm{B},E_\mathrm{kin}) 
    &\xrightarrow{\theta^{-1}_{f,\text{sim}}(\theta_i; B,U_\mathrm{B},E_\mathrm{kin})}&
    \theta_i\\
    \varphi_f(B,U_\mathrm{B},E_\mathrm{kin}) 
    &\xrightarrow{\varphi^{-1}_{f,\text{sim}}(\varphi_i;\theta_i, B,U_\mathrm{B},E_\mathrm{kin})}&
    \varphi_i.
\end{eqnarray}
where $\theta_{f,\text{sim}}$ and $\varphi_{f,\text{sim}}$ are the functions obtained from our trajectory simulation. We note that $\theta_{f,\text{sim}}$ does not depend on the initial $\varphi_i$ angles due to the axial symmetry of the solenoid magnetic field. In practice, we calculate correction matrices of $\theta_{f,\text{sim}}$ and $\Delta\varphi_{\text{sim}}$ on a dense grid of $\{\theta_i, E_\mathrm{kin}, B\}$ for each $U_\mathrm{B}$. Here, $\Delta\varphi = \varphi_f-\varphi_i$. These matrices are then linearly interpolated. The intensity of the corrected ARPES cube at each $(\varphi_i,\theta_i, E_\mathrm{kin})$ is obtained by interpolating the intensity of the measured cube around the corresponding $(\varphi_f, \theta_f, E_\mathrm{kin})$ obtained from the correction matrices.

We perform the corrections on the complete ARPES cubes. Slices through the corrected cubes are shown in Fig.~\ref{Fig:ARPES}(d,j). They demonstrate the high fidelity of the angle correction procedure. Some deviations are visible close to the low-energy cut-off (-0.5\,eV at $0^\circ$), where the small kinetic energies lead to the most substantial field aberrations of the electron trajectories. Small asymmetric distortions of the Fermi surface are visible particularly for $U_\mathrm{B}=20$\,V (Fig.~\ref{Fig:ARPES}(j)). These deviations are attributed to a slight offset between the beam spot and the magnetic-field center. This highlights the importance of precise sample and beam positioning in magneto-ARPES experiments. Slight distortions at zero field are related to inhomogeneous electric fields from the work function difference between \BS~and its immediate surroundings (Cu, Al). They are common in laser-ARPES and imaging performance improves with larger bias \cite{pfau_2020}.

To quantify the reconstruction accuracy, we show momentum-distribution curves (MDCs) from the corrected Fermi surfaces in Fig.~\ref{Fig:ARPES}(e,k). The deviations of the angle $\theta_\mathrm{F}$ that corresponds to the Fermi wave vector remain within $\pm1^\circ$ for all fields and biases (Fig.~\ref{Fig:ARPES}(f,l)). For comparison, the intrinsic linewidth at zero field is $\Delta \theta_0 = 0.8^\circ$ ($1.1^\circ$) for $U_\mathrm{B}=80$\,V (20\,V) and therefore of the same magnitude. $\Delta \theta_\mathrm{F}$ is dominated by the asymmetric Fermi surface distortions discussed above, because it has no systematic field dependence and the opposite sign along $k_x$ and $k_y$.

% -----------------------------------------------------------
% Figure: Dirac gap
% -----------------------------------------------------------
\begin{figure}
\includegraphics[width=\columnwidth]{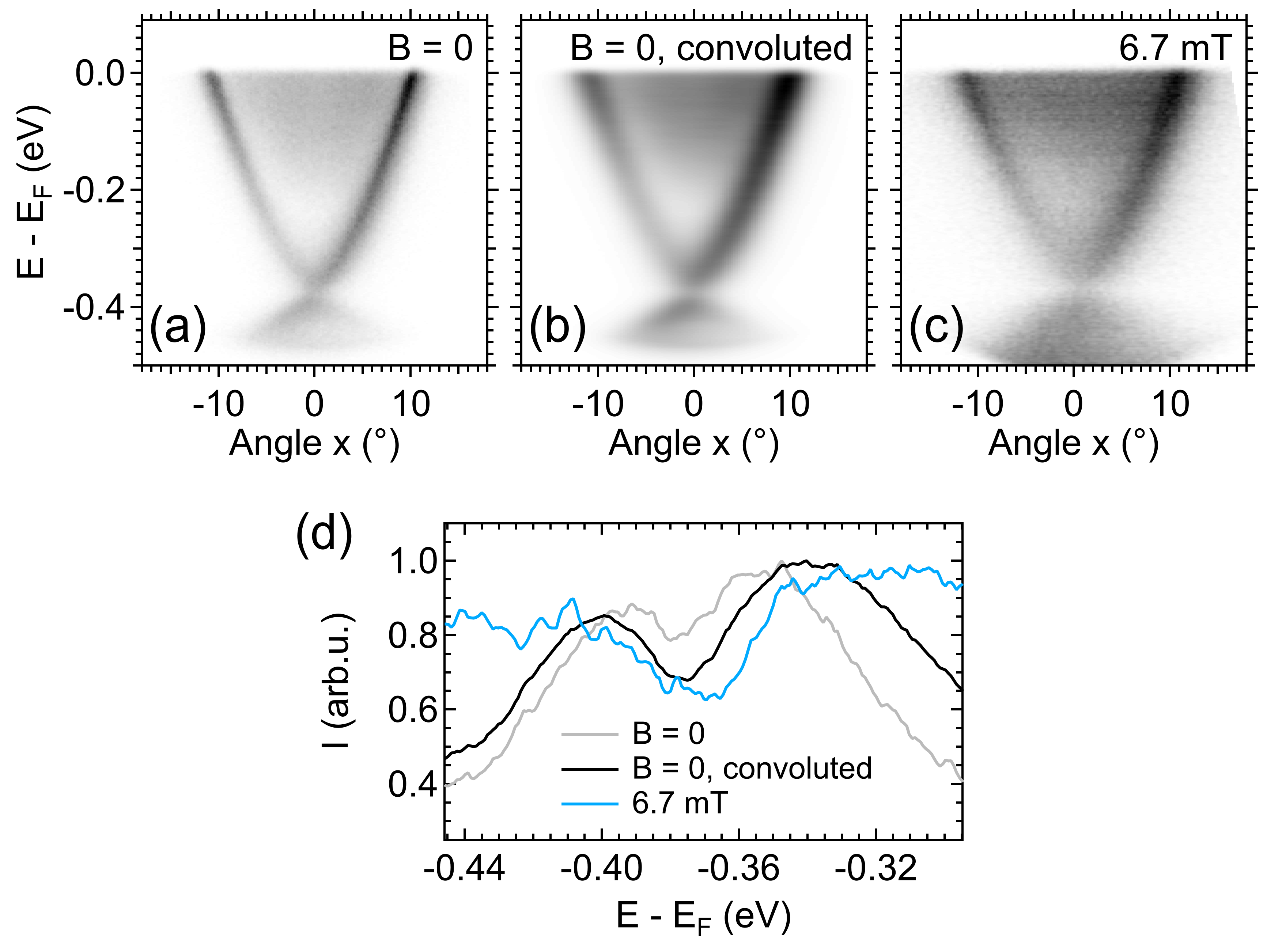}
\caption{
Analysis of the Dirac point. (a) Spectrum through the Dirac point at zero field. (b) Same spectrum, but extracted from the ARPES cube after convolution with a Gaussian of width $1.2^\circ$ in both angle directions. The width corresponds to the field-broadening observed at 6.7\,mT (\cf~Fig.~\ref{Fig:ARPES}(b)). (c) Spectrum at 6.7\,mT. (d) Energy-distribution curves (EDCs) through the Dirac point for all three spectra. The dip in intensity around -0.38\,eV corresponds to the location of the Dirac point. Broadening leads to an apparent opening of a gap.  
}
\label{Fig:gap}
\end{figure}

Angular broadening becomes visible in the spectra at the largest fields. We quantify the broadening at $k_\mathrm{F}$ from the angle corrected spectra in Fig.~\ref{Fig:ARPES}(d,j) using $\Delta \theta = \sqrt{\Delta \theta_B^2 - \Delta \theta_0^2}$. The results in Fig.~\ref{Fig:ARPES}(b,h) show that the field broadening is less than $2\Delta \theta_0$ and in agreement with our simulations. Broadening is therefore dominated by the different fields that the electrons from different locations of the sample experience. Possible additional sources such as a larger virtual spot\cite{huang_2023,Li_2025} appear negligible. The high resolution of our magneto-ARPES data is a consequence of the small 20\,$\mu$m beam spot and precise sample and beam positioning. 

A magnetic field is expected to open a gap at the Dirac point, which seems to be present in the spectra in Fig.~\ref{Fig:ARPES}(d). However, it is most likely an artifact related to momentum broadening as discussed in previous magneto-ARPES studies \cite{huang_2023}. We demonstrate this effect in Fig.~\ref{Fig:gap}, where we compare the data at 6.7\,mT with the zero-field data convoluted with the field-broadening. Larger fields and/or a smaller spot size are required to unambiguously observe the gap opening.

We use the measurements at $U_\mathrm{B}=20$\,V to explore imaging across the focal point at approximately 7\,mT. Close to the critical field, spectra are contracted to an extent that no angle distribution can be resolved (not shown). For a larger field (-8.6\,mT), contraction reduces again and the angles move to the opposite geometric quadrant confirming our simulations. While broadening is the largest at -8.6\,mT, it is predicted to reduce substantially (Fig.~\ref{Fig:ARPES}(h)) once $\theta_f$ reaches its next maximum at 11\,mT (Fig.~\ref{Fig:ARPES}(g)).

%%%%%%%%%%%%%%%%%%%%%%%%%%%%%%%%%%%%%%%%%%%%%%%%%%%%%%%%%%%%%%
\section{Circular dichroism in magneto-ARPES of B\lowercase{i}$_2$S\lowercase{e}$_3$}
%%%%%%%%%%%%%%%%%%%%%%%%%%%%%%%%%%%%%%%%%%%%%%%%%%%%%%%%%%%%%%

% -----------------------------------------------------------
% Figure: CD
% -----------------------------------------------------------
\begin{figure}
\includegraphics[width=\columnwidth]{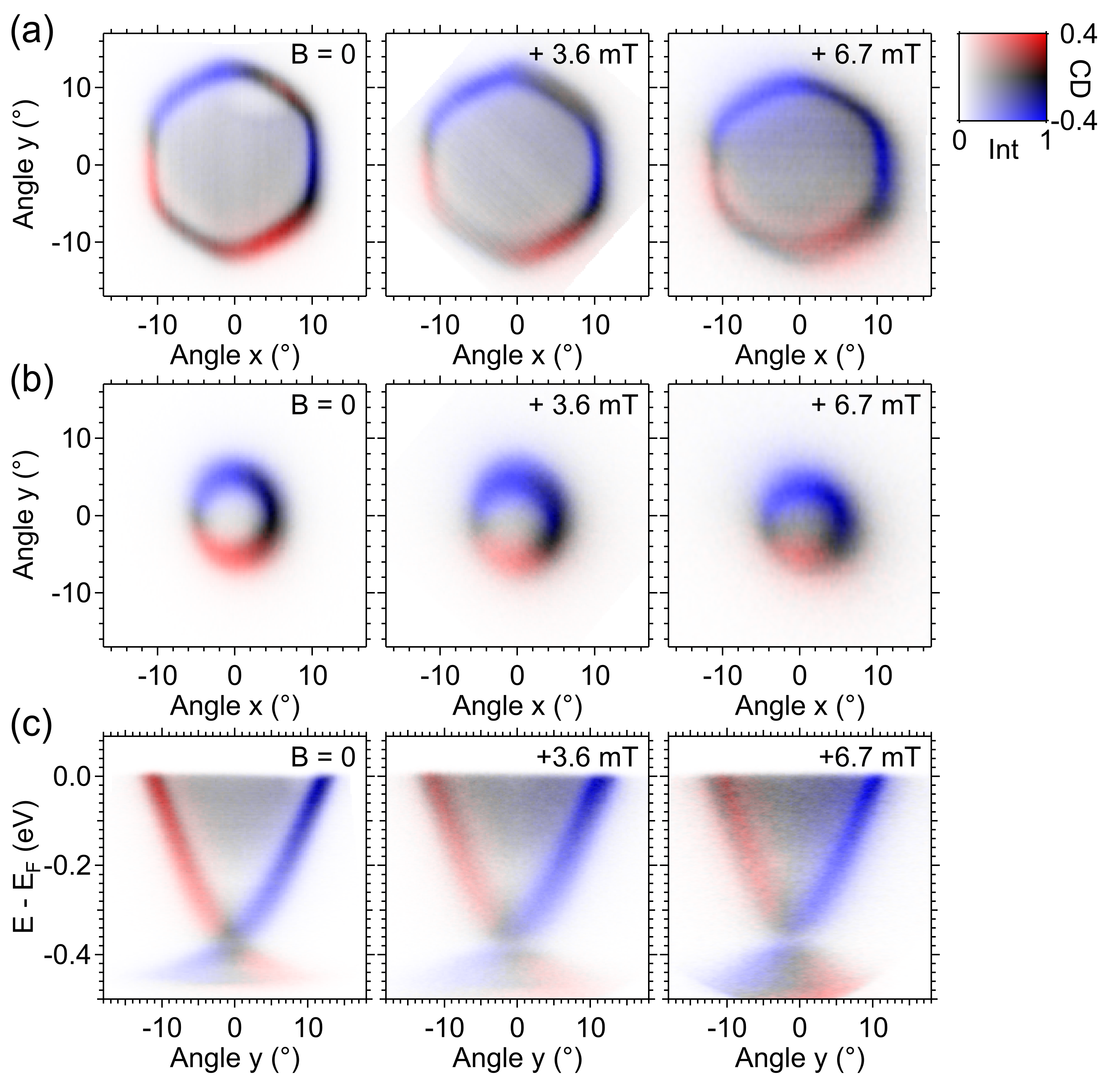}
\caption{
Field-dependent CD in \BS~of (a) the Fermi surface, (b) an equi-energy surface at -0.25\,eV slightly above the Dirac point, and (c) an energy-angle cut through the Dirac point. All spectra are extracted from angle-corrected magneto-ARPES cubes. The 2D color scale represents the difference (sum) of spectra taken with left and right circularly polarized light along the CD (Int) axis.
}
\label{Fig:CD}
\end{figure}

The spin texture described by the Hamiltonian in Eq.~(\ref{eqn:H}) is coupled to an equivalent texture of the orbital angular momentum (OAM) \cite{zhang_2013}, which can be accessed through circular dichroism (CD) in ARPES \cite{wang_2011,park_2012}. In general, CD in magneto-ARPES can be a powerful tool to detect field-induced changes of the spin structure in materials with spin-orbit coupling. 

Close to the Dirac point in \BS, the OAM has an in-plane helical structure while it acquires an out-of-plane component for larger momenta close to the Fermi level. Field-induced changes are only expected in the gapped region of the Dirac cone within $g\mu_\mathrm{B} B$, where an out-of-plane spin alignment develops, with corresponding signatures in the OAM.

Fig.~\ref{Fig:CD} summarizes our CD measurements as a function of the field. We observe the expected helical texture of the OAM close to the Dirac point (Fig.~\ref{Fig:CD}(b,c)). The Fermi surface shows a more complex pattern formed by a superposition of in- and out-of-plane OAM components, as described in previous ARPES measurements \cite{wang_2011}. They are a consequence of the $\alpha$ and $\gamma$-terms in the Hamiltonian (\ref{eqn:H}), respectively. The overall CD pattern does not change in field, which exemplifies the excellent performance of our angle-correction algorithm. The full energy and angle dependence of the electron trajectories needs to be taken into account to successfully recover the OAM structure. We cannot experimentally resolve changes at the Dirac point itself due to the small field range and angle broadening.

%%%%%%%%%%%%%%%%%%%%%%%%%%%%%%%%%%%%%%%%%%%%%%%%%%%%%%%%%%%%%%
\section{Discussion}
%%%%%%%%%%%%%%%%%%%%%%%%%%%%%%%%%%%%%%%%%%%%%%%%%%%%%%%%%%%%%%

We demonstrate that spectral distortions in magneto-ARPES can be corrected with high fidelity in post-processing using detailed electron trajectory simulations. The magnetic dipole field acts as an electron lens and focal points recur at specific magnetic-field values, where all trajectories are straight at the analyzer entrance. Importantly, we show that magneto-ARPES can be performed in between these focal points. 

Both a bias voltage and higher photon energies shift the focal points in field, which can be selected to optimize imaging at magnetic fields relevant for the sample physics. Any solenoid setup aiming to increase the field range of magneto-ARPES will need to consider focal points and employ complete angle and energy-dependent corrections even at high photon energies.

We demonstrate that a small spot size is the key to performing magneto-ARPES with high momentum resolution. Beyond the micro-ARPES regime, momentum broadening quickly becomes unsustainable. Strong angle contraction around the focal points increases broadening substantially and prevents high-resolution measurements in these field regions regardless of the spot size. Importantly, momentum resolution is recovered between focal points, which will be instrumental to perform ARPES at larger fields in the future. While larger photon energies provide an advantage at small fields, we show that they face the same general scaling laws with excessive broadening close to the focal points.

Previous studies used ferromagnetic cores to enhance the field magnitude \cite{ryu_2023} and higher multi-pole configurations to shape the vector field \cite{Li_2025}. Combining these approaches with the insights from our high-resolution measurements and detailed simulations will lay the foundation for future developments in magneto-ARPES.

%%%%%%%%%%%%%%%%%%%%%%%%%%%%%%%%%%%%%%%%%%%%%%%%%%%%%%%%%%%%%%
% acknowledgments
%%%%%%%%%%%%%%%%%%%%%%%%%%%%%%%%%%%%%%%%%%%%%%%%%%%%%%%%%%%%%%

\begin{acknowledgments}
We are very grateful for valuable discussions with Patrick Kirchmann, Yu He, and Jonathan Sobota. This work is supported by the U.S. Department of Energy (DOE), Office of Science, Office of Basic Energy Sciences, Materials Sciences and Engineering Division, under Award Number DE-SC0024135. C.X.L and D.Z.J also acknowledge the support from the ICDS Faculty Upskilling award ICDS\_UP25\_028016 from Penn State's Institute for Computational and Data Sciences (RRID:SCR 025154). H.T.L and C.X.L were partially supported by the Rising Researcher award ICDS RR25 27708 from Penn State’s Institute for Computational \& Data Sciences (RRID:SCR 025154). The sample growth effort carried out in this work was supported by the DOE under grant DE-SC0019068. 
\end{acknowledgments}

%%%%%%%%%%%%%%%%%%%%%%%%%%%%%%%%%%%%%%%%%%%%%%%%%%%%%%%%%%%%%%
% Bibliography
%%%%%%%%%%%%%%%%%%%%%%%%%%%%%%%%%%%%%%%%%%%%%%%%%%%%%%%%%%%%%%
\nocite{*}
\bibliography{main}

\end{document}